\documentstyle [11pt]{article}

\begin{document}

\begin{center}
{\huge \bf Two-boson realizations of the Higgs algebra and some applications}
\end{center}

\vspace{2mm}

\begin{center}
Dong Ruan
\end{center}
{\it 
Department of Physics, Tsinghua University,
      Beijing 100084, China, 
Center of Theoretical Nuclear Physics,
      National Laboratory of Heavy Ion Accelerator,
      Lanzhou, 730000, China, and
Key Laboratory for Quantum Information and
    Measurements of Ministry of Education,
    Tsinghua University, Beijing 100084, China
}

\vspace{3mm}

\begin{abstract}
In this paper two kinds of two-boson realizations of the Higgs
algebra are obtained by generalizing the well known
Jordan-Schwinger realizations of the SU(2) and SU(1,1) algebras.
In each kind, an unitary realization and two nonunitary
realizations, together with the properties of their respective acting
spaces are discussed in detail. Furthermore, similarity transformations,
which connect the nonunitary realizations with the unitary ones,
are gained by solving the corresponding unitarization equations.
As applications, the dynamical symmetry of the Kepler system in a
two-dimensional curved space is studied and the phase operators of
the Higgs algebra is constructed.
\end{abstract}

\vspace{4mm}

\section{Introduction}
\label{1}

In recent years, the polynomial angular momentum algebra (PAMA)
and its increasing applications in quantum problems have been the
focus of very active research. This kind of PAMA, spanned by three
elements ${\cal J}_{\mu}$ ($\mu = +$, $-$, $3$), has a coset
structure $h+ v$, \cite{rocek} where $h$ is an ordinary Lie
algebra U(1) generated by ${\cal J}_3$; the remaining two elements
${\cal J}_+$, ${\cal J}_-$ $\in v$ transform according to a
representation of U(1), and their commutator yields a polynomial
function of order $n$ in the operator ${\cal J}_3 \in$ U(1), i.e.,
\begin{equation}
    [ {\cal J}_3, \hspace{1mm} {\cal J}_{\pm} ] = \pm {\cal J}_{\pm},
\hspace{4mm}
    [ {\cal J}_+, \hspace{1mm} {\cal J}_-] =
        \sum_{i=0}^{n} C_i ( {\cal J}_3 )^i ,
\label{pama}
\end{equation}
where the coefficients $ C_i $ ($i=0$, 1, ..., $n$) are real constants.
When $C_1=2$ (or $-2$) and $C_0=C_j=0$ ($j \geq 2$), Eq. (\ref{pama})
goes back to the commutation relations satisfied by
the angular momentum algebra SU(2) (or its non-compact type
SU(1,1)). \cite{bl} Hence, the PAMA can be viewed as a type of polynomial
deformation of SU(2) (or SU(1,1)), or a type of nonlinear extension of U(1).

The first special case of the PAMA is the so-called Higgs algebra,
which, here denoted by ${\cal H}$, was used by Higgs \cite{higgs}
to establish the existence of additional symmetries for the
isotropic oscillator and Kepler potentials in a two-dimensional
curved space. Later, Zhedanov \cite{zhedanov} presented a
connection between the Higgs algebra ${\cal H}$ and the quantum
group SU$_q$(2). \cite{jimbo} Daskaloyannis \cite{dask} and
Bonatsos {\it et al.} \cite{bdk,bkd} discussed the PAMA by means
of the generalized deformed oscillator, respectively, and Quesne
\cite{quesne1} related it to the generalized deformed parafermion.
Junker {\it et al.} \cite{jr} constructed the (nonlinear) coherent
states of ${\cal H}$ for the conditionally exactly solvable model
with the radial potential of harmonic oscillator, and Kumar {\it
et al.} \cite{sbjps} did for the quadrilinear boson Hamiltonian
describing four-photon process and showed \cite{sbj} that the PAMA
of order ($n_1 + n_2 + 1$) may be constructed by combining two
given mutually commuting PAMAs with their respective orders being
$n_1$ and $n_2$. Recently, Beckers {\it et al.} \cite{bbd} and
Debergh \cite{debergh-1,debergh-2} realized ${\cal H}$, which is
seen as a spectrum generating algebra in their method, by
single-variable differential operators in the study of
(quasi-)exactly solvable problems, and also construct a special
unitary two-boson realization to study the Karassiov-Klimov
Hamiltonian in the quantum optics. Ruan {\it et al.} \cite{rjr}
studied indecomposable representations of the PAMA of quadratic
type, and then from these representations obtained its
inhomogeneous boson realizations. In the present work we will
study in detail for ${\cal H}$ two-boson realizations, which are
analogous to the well known Jordan-Schwinger realizations of the
SU(2) and SU(1,1) algebras, \cite{bl} and some applications.

\par
This paper is arranged as follows. In Sec. \ref{2}, some
elementary results of the Jordan-Schwinger realizations of SU(2)
and SU(1,1) and of the irreducible unitary representations of
${\cal H}$ are briefly reviewed, respectively. In Sec. \ref{3},
two kinds of two-boson realizations of ${\cal H}$ are studied in
detail, such as the unitary realizations, the nonunitary realizations,
and their respective acting spaces.
In Sec. \ref{4}, we first discuss generally the unitarization
equations satisfied by the nonunitary realizations, then calculate
the explicit expressions for the corresponding similarity
transformations, which may relate the nonunitary
realizations to the unitary ones. In Sec. \ref{5}, as applications,
by making use of the results obtained in Sec. \ref{3},
the dynamical symmetry of the Kepler system in the
two-dimensional curved space is studied and the phase
operators of ${\cal H}$ are constructed. A simple discussion is given in
the final section.

\section{NOTATIONS AND SOME ELEMENTARY RESULTS}
\label{2}

In this section, some elementary results, along with notations, to be used later
are briefly reviewed, such as the standard Jordan-Schwinger realizations
of the SU(2) and SU(1,1) algebras, the irreducible unitary representations
of the Higgs algebra ${\cal H}$, and so on.

\subsection{The Jordan-Schwinger realizations of SU(2) and SU(1,1)}

Denote three generators of SU(2) and its non-compact type SU(1,1)
by \{$J_+$, $J_-$, $J_3$\}, then their commutation relations may
be written in a compact form
\begin{equation}
    [ J_+, \hspace{1mm} J_-] = 2 \lambda J_3  ,  \hspace{4mm}
    [ J_3, \hspace{1mm} J_{\pm} ] = \pm J_{\pm},
\label{am}
\end{equation}
where $\lambda = 1$ for SU(2) and $\lambda = -1$ for SU(1,1).

Schwinger \cite{schwinger} found that three components of the
angular momentum ${\bf J}$ may be described by means of the
occupation number representation of the two-dimensional isotropic
harmonic oscillator. In terms of the famous Jordan-Schwinger
mapping, \cite{schwinger} the generators of SU(2) and SU(1,1) may
be respectively realized by two pairs of mutually commuting boson
operators \{$a_i, \, a^+_i| \; i=1, \, 2 $\} (the annihilation
operators $a_i$ are adjoint to the creation operators $a^+_i$,
i.e., $a_i = (a^+_i)^\dagger$, $a^+_i = (a_i)^\dagger$) as
\begin{eqnarray}
\begin{array}{l}
    J_+ = a^+_1 a_2, \\
    J_- = a_1 a^+_2 , \\
    J_3 = {1 \over 2}( \hat{n}_1 - \hat{n}_2 )
\label{su2}
\end{array}
\end{eqnarray}
for SU(2), and
\begin{eqnarray}
\begin{array}{l}
    J_+ = a^+_1 a^+_2, \\
    J_- = a_1 a_2, \\
    J_3 = {1 \over 2}( \hat{n}_1 + \hat{n}_2 + 1 )
\label{su11}
\end{array}
\end{eqnarray}
for SU(1,1), where $\hat{n}_i$ $\equiv$ $a^+_i a_i$ ($i=1$, 2) are the corresponding
particle number operators, which, together with the boson operators \{$a_i, \, a^+_i$\},
satisfy the commutation relations
\begin{eqnarray}
\begin{array}{l}
    [ a_i, \; a^+_j ] = \delta_{ij},  \cr
    [ \hat{n}_i, \mbox{} a^+_j ] = \delta_{ij} a^+_j  , \cr
    [ \hat{n}_i, \mbox{} a_j ] = - \delta_{ij} a_j.
\end{array}
\label{an-cr}
\end{eqnarray}
Furthermore, the complete set of basis vectors of Fock space,
$ {\cal F} \equiv \{| n_1 n_2 \rangle| \: n_1,\, n_2=0$, 1, 2, ...\}, may be
constructed from the vacuum state $| 00 \rangle$ of
the two-dimensional harmonic oscillator by using the definition
\begin{eqnarray}
   | n_1 n_2 \rangle = \frac{ (a^+_1)^{n_1} (a^+_2)^{n_2} }
    { \sqrt{n_1 ! n_2 !} } | 00 \rangle.
\label{n12-state}
\end{eqnarray}
In fact, these vectors are the common normalized eigenvectors of $\hat{n}_1$ and
$\hat{n}_2$ belonging to eigenvalues $n_1$ and $n_2$ respectively, i.e.,
\begin{equation}
  \hat{n}_i | ... n_i ... \rangle = n_i | ... n_i ... \rangle,
    \hspace{4mm}
    i=1,\: 2,
\label{n12-eigen}
\end{equation}
and satisfy
\begin{eqnarray}
\begin{array}{l}
    a_i | ... n_i ... \rangle =
        \sqrt{n} | ... n_i -1... \rangle, \cr
    a^+_i | ... n_i ... \rangle =
        \sqrt{n+1} | ... n_i +1 ... \rangle.
\end{array}
\label{aa+n}
\end{eqnarray}

Correspondingly, the common eigenvectors $| j m \rangle$ of the angular
momentum operators ${\bf J}^2$ and $J_3$ may also be expressed in
the Jordan-Schwinger representation as
\begin{eqnarray}
   | j m \rangle = \frac{ (a^+_1)^{j+m} (a^+_2)^{j-m} }
    { \sqrt{(j+m)! (j-m)!} } | 00 \rangle.
\label{amb-a}
\end{eqnarray}
Comparison between Eq. (\ref{amb-a}) and Eq. (\ref{n12-state})
leads immediately to
\begin{eqnarray}
   \hat{n}_i | j m \rangle = [j- (-1)^i m ] | j m  \rangle,
    \hspace{4mm}
    i=1,\: 2,
\label{n12-am-b}
\end{eqnarray}
that is, the quantum numbers $n_1$ and $n_2$ are related to $j$ and $m$ by the equations
$n_1 = j+ m$ and $n_2 = j- m$.

The other useful methods that realize the ordinary Lie algebras by
bosons may be found in Refs. \cite{km,fradkin}

\subsection{The Higgs algebra ${\cal H}$ and its irreducible unitary representation}
\label{2-2} Taking $C_2 = C_j = 0$ ($j>3$) in Eq. (\ref{pama}), it
follows that the three generators \{${\cal J}_{\pm}$, ${\cal
J}_3$\} of the Higgs algebra ${\cal H}$ satisfy the following
commutation relations
\begin{eqnarray}
    [ {\cal J}_3, \hspace{1mm} {\cal J}_{\pm} ] = \pm {\cal J}_{\pm} ,  \hspace{4mm}
    [ {\cal J}_+, \hspace{1mm} {\cal J}_-] = C_1 {\cal J}_3 + C_3 {\cal J}^3_3.
\label{higgs}
\end{eqnarray}
In analogy with SU(2), \cite{bl} the Casimir invariant of ${\cal
H}$ reads
\begin{equation}
{\cal C} = \frac12 ( {\cal J}_+ {\cal J}_- + {\cal J}_- {\cal J}_+) +
        \left ( \frac12 C_1 + \frac14 C_3 \right) {\cal J}_3^2
             + \frac14 C_3 {\cal J}_3^4 ,
\label{casimir}
\end{equation}
which commutes with the three generators of ${\cal H}$,
i.e.,
\begin{equation}
 [{\cal C}, \mbox{} {\cal J}_{\pm} ] =
 [{\cal C}, \mbox{} {\cal J}_3 ] = 0.
\end{equation}
It is worthy of reminding the readers that the constant $C_1$ in
Eq. (\ref{higgs}) is remained for convenience though it may become
some fixed real number, say $q$, by rescaling the generators,
${\cal J}_{\pm} \rightarrow \sqrt{q / C_1} {\cal J}_{\pm}$.

Making use of the parallel treatment of angular momentum in
quantum mechanics, \cite{bl} it is not difficult to obtain the
following unitary representation of ${\cal H}$ in the common
eigenvectors $| \tilde{j} \tilde{m} \rangle$ of the elements
\{${\cal C}$, ${\cal J}_3$\}, with $\tilde{j}$ and $\tilde{m}$
labelling the eigenvalues of ${\cal C}$ and ${\cal J}_3$,
respectively, \cite{bbd,ruan}
\begin{equation}
\begin{array}{rl}
\langle \tilde{j} \tilde{m}+1 | {\cal J}_+ | \tilde{j} \tilde{m} \rangle =
    &  \sqrt{ \frac12 C_1 [ \tilde{j}(\tilde{j}+1)- \tilde{m}(\tilde{m}+1)]
    + \frac14 C_3[\tilde{j}^2 (\tilde{j}+1)^2 - \tilde{m}^2 (\tilde{m}+1)^2 ]}, \\
\langle \tilde{j} \tilde{m}-1 | {\cal J}_- | \tilde{j} \tilde{m} \rangle =
    &  \sqrt{ \frac12 C_1[\tilde{j}(\tilde{j}+1)- \tilde{m}(\tilde{m}-1)]
    + \frac14 C_3[\tilde{j}^2 (\tilde{j}+1)^2 - \tilde{m}^2 (\tilde{m}-1)^2 ]}, \\
\langle \tilde{j} \tilde{m} | {\cal J}_3 | \tilde{j} \tilde{m} \rangle =
    &  \tilde{m},   \\
\langle \tilde{j} \tilde{m} | {\cal C} | \tilde{j} \tilde{m} \rangle =
    &  \frac12 C_1 \tilde{j} (\tilde{j}+1) + \frac14 C_3 \tilde{j}^2 (\tilde{j}+1)^2.
\end{array}
\label{urep}
\end{equation}
Here we have adopted the same phase factor as the Condon-Shortley
convention of SU(2) so that the matrix elements of ${\cal
J}_{\pm}$ are real. In Eq. (\ref{urep}), $\tilde{j}$ may take
half-integers, i.e., 0, 1/2, 1, 3/2,..., and for the finite dimensional
representation with a fixed $\tilde{j}$, the values that $m$ may
take, being a part of \{$-\tilde{j}$, $-\tilde{j}+ 1$,...,
$\tilde{j}$\}, are different for different $C_1$'s and $C_3$'s.
\cite{ruan}

\section{TWO KINDS OF TWO-BOSON REALIZATIONS OF ${\cal H}$}
\label{3} In this section, we will study two kinds of two-boson
realizations of ${\cal H}$, which are analogous to the
Jordan-Schwinger realizations of SU(2) and SU(1,1), respectively.

\subsection{The first kind of realizations }
\label{3-1}

The Jordan-Schwinger realization (\ref{su2}) of SU(2) reminds us that
the first kind of two-boson realizations of ${\cal H}$ may be chosen in
the following form
\begin{eqnarray}
\begin{array}{rl}
    \dot{B}^{(k,l)} ({\cal J}_+) & = \dot{f} (\hat{n}_1,\hat{n}_2) (a_1^+)^k a_2^l,  \\
    \dot{B}^{(k,l)} ({\cal J}_-) & = a_1^k (a_2^+)^l \dot{g} (\hat{n}_1,\hat{n}_2), \\
    \dot{B}^{(k,l)} ({\cal J}_3) & = \dot{h} (\hat{n}_1,\hat{n}_2),
\end{array}
\label{h-su2-jj}
\end{eqnarray}
where $k$ and $l$ are positive integers,
$\dot{f} (\hat{n}_1,\hat{n}_2)$, $\dot{g} (\hat{n}_1,\hat{n}_2)$ and
$\dot{h} (\hat{n}_1,\hat{n}_2)$,
being the operator functions of $\hat{n}_1$ and $\hat{n}_2$, have to be determined
by the commutation relations (\ref{higgs}) of ${\cal H}$.
For a fixed $(k,l)$, the action of $\dot{B}^{(k,l)} ({\cal J}_{\pm})$ on some basis vector
$| n_1 n_2 \rangle$ of the boson Fock space ${\cal F}$ gives another
basis vector $| n_1 \pm k, n_2 \mp l\rangle$.

The first equation of Eq. (\ref{higgs}) requires that $\dot{h} (\hat{n}_1,\hat{n}_2)$
satisfies the simple two-variable difference equation
\begin{equation}
    \dot{h} (\hat{n}_1,\hat{n}_2) - \dot{h} (\hat{n}_1 - k ,\hat{n}_2 + l) =1.
\label{h-equ}
\end{equation}
Its solution reads
\begin{equation}
    \dot{h} (\hat{n}_1,\hat{n}_2) =
        \frac{\hat{n}_1}{2 k} - \frac{\hat{n}_2}{2 l} + \alpha,
\label{h-solu}
\end{equation}
here $\alpha$, being a real constant, needs further determining by considering
the irreducible representation of ${\cal H}$ given in Section \ref{2-2}.
Equation (\ref{h-solu}) clearly shows that
the two-boson realization (\ref{h-su2-jj}) can not be reduced to
the single-boson case by setting $k=0$ or $l=0$ because of singularity.

Substituting Eq. (\ref{h-solu}) into Eq. (\ref{h-su2-jj}), thus,
satisfaction of the second equation of Eq. (\ref{higgs}) requires
that $\dot{f} (\hat{n}_1,\hat{n}_2) \dot{g} (\hat{n}_1,\hat{n}_2)$
satisfies the following two-variable difference equation
\begin{equation}
\begin{array}{l}
    \left [ \prod\limits_{i=1}^{k} (\hat{n}_1 - i +1) \right ]
    \left [ \prod\limits_{i=1}^{l} (\hat{n}_2 + i) \right ]
     \dot{f}^{(k,l)} (\hat{n}_1,\hat{n}_2) \dot{g}^{(k,l)}(\hat{n}_1,\hat{n}_2)       \\
    - \left [ \prod\limits_{i=1}^{k} (\hat{n}_1 + i) \right ]
    \left [ \prod\limits_{i=1}^{l} (\hat{n}_2 - i + 1) \right ]
     \dot{f}^{(k,l)}(\hat{n}_1 + k ,\hat{n}_2 - l)
        \dot{g}^{(k,l)}(\hat{n}_1 + k ,\hat{n}_2 - l) \\
    = C_1 \left ( \frac{\hat{n}_1}{2 k} - \frac{\hat{n}_2}{2 l} + \alpha \right )
    + C_3  \left ( \frac{\hat{n}_1}{2 k} - \frac{\hat{n}_2}{2 l} + \alpha \right )^3.
\label{fg-su2}
\end{array}
\end{equation}
In the process of obtaining the above equation, we have used the fundamental
relations
\begin{equation}
\begin{array}{rl}
    a^k_i f (...\hat{n}_i...) = & f (...,\hat{n}_i + k,...) a^k_i,
\hspace{4mm} i=1, \:2, \\
    (a^+_i)^k f (...\hat{n}_i...) = & f (...,\hat{n}_i - k,...) (a^+_i)^k,
\end{array}
\end{equation}
which follow from Eq. (\ref{an-cr}) for any function $f (...\hat{n}_i...)$.

Note that Eq. (\ref{fg-su2}) only fixes the product $\dot{f}
(\hat{n}_1,\hat{n}_2) \dot{g} (\hat{n}_1,\hat{n}_2)$. Different
choices of the two functions, as well as the constant $\alpha$, may
produce a variety of realizations for ${\cal H}$. However, it is
very difficult to obtain the general solutions of Eq. (\ref{fg-su2})
for arbitrary $k$ and $l$. Below will study in more detail the special case of
$(k,l)=(1,1)$, and give directly the results of the case of $(k,l)=(2,2)$.

\vspace{2mm}
{\bf 1. The $(1,1)$ case.}

Inserting $k=l=1$ into Eq. (\ref{fg-su2}) and solving it, we may obtain the following
two solutions
\begin{eqnarray}
\begin{array}{rl}
    \dot{f}^{(1,1)}_1 (\hat{n}_1,\hat{n}_2) \dot{g}^{(1,1)}_1 (\hat{n}_1,\hat{n}_2) = &
         \frac{1}{8 \hat{n}_1} (\hat{n}_1 + 2 \alpha )
          \{ 4 C_1 + C_3 [ \hat{n}_1 (\hat{n}_1 + 4 \alpha )  \\
    {}  &   + (\hat{n}_2 + 1)^2 + (2 \alpha +1)(2 \alpha -1)] \}
\end{array}
\label{solu-1}
\end{eqnarray}
and
\begin{eqnarray}
\begin{array}{rl}
    \dot{f}^{(1,1)}_2 (\hat{n}_1,\hat{n}_2) \dot{g}^{(1,1)}_2 (\hat{n}_1,\hat{n}_2) = &
         \frac{1}{8 (\hat{n}_2 + 1)} (\hat{n}_2 - 2 \alpha +1 )
          \{ 4 C_1 + C_3 [ \hat{n}_1^2  \\
    {}  &   + \hat{n}_2 ( \hat{n}_2 - 4 \alpha +2 ) + 4 \alpha (\alpha -1)] \}.
\end{array}
\label{solu-2}
\end{eqnarray}
From them we have some freedom in the choice of the functions
$\dot{f}^{(1,1)}_i (\hat{n}_1,\hat{n}_2)$ ($i=1$, 2) and
$\dot{g}^{(1,1)}_i(\hat{n}_1,\hat{n}_2)$. However here we need
only consider the first solution (\ref{solu-1}) because of the
symmetry between the solutions (\ref{solu-1}) and (\ref{solu-2})
$$
    \hat{n}_1 \leftrightarrow \hat{n}_2 +1
\hspace{2mm} \mbox{and} \hspace{2mm}
    \alpha \leftrightarrow - \alpha.
$$

(1) If the unitary relations need satisfying, i.e.,
\begin{equation}
 \dot{B}^{(1,1)} ({\cal J}_{\pm}) = \left ( \dot{B}^{(1,1)} ({\cal J}_{\mp})
    \right )^{\dagger},
\label{u-c}
\end{equation}
($\dot{B}^{(1,1)} ({\cal J}_3)$ is already hermitian), which lead to
$\dot{f}^{(1,1)}_1 (\hat{n}_1,\hat{n}_2) = \dot{g}^{(1,1)}_1 (\hat{n}_1,\hat{n}_2)$,
then solving Eq. (\ref{solu-1}) and substituting the expression of
$\dot{f}^{(1,1)}_1 (\hat{n}_1,\hat{n}_2)$ into Eq. (\ref{h-su2-jj}),
we may obtain
\begin{eqnarray}
\begin{array}{rl}
\dot{B}^{(1,1)}_1 ({\cal J}_+) = & \{ \frac{1}{8 \hat{n}_1} (\hat{n}_1 + 2 \alpha )
          \{ 4 C_1 + C_3 [ \hat{n}_1 (\hat{n}_1 + 4 \alpha )  \\
    {}  &   + (\hat{n}_2 + 1)^2 + (2 \alpha +1)(2 \alpha -1)] \} \}^{1/2}
        a_1^+ a_2,    \\
\dot{B}^{(1,1)}_1 ({\cal J}_-) = & a_1 a^+_2 \{
        \frac{1}{8 \hat{n}_1} (\hat{n}_1 + 2 \alpha )
          \{ 4 C_1 + C_3 [ \hat{n}_1 (\hat{n}_1 + 4 \alpha )  \\
    {}  &   + (\hat{n}_2 + 1)^2 + (2 \alpha +1)(2 \alpha -1)] \} \}^{1/2},    \\
\dot{B}^{(1,1)}_1 ({\cal J}_3)  = & \frac{1}{2} (\hat{n}_1 - \hat{n}_2) + \alpha.
\end{array}
\label{su2-11-1}
\end{eqnarray}
It can be easily checked that the realization (\ref{su2-11-1}) satisfies
Eq. (\ref{higgs}) for arbitrary $\alpha$.

Inserting Eq. (\ref{su2-11-1}) into Eq. (\ref{casimir}), the Casimir invariant ${\cal C}$
of ${\cal H}$ may be expressed in terms of the boson number operators $n_1$ and $n_2$ as
\begin{equation}
    {\cal C} = \frac{1}{64} (\hat{N} + 2 \alpha) (\hat{N} + 2 \alpha + 2)
    [8 C_1 + C_3  (\hat{N} + 2 \alpha) (\hat{N} + 2 \alpha + 2) ],
\label{h-su2-casimir}
\end{equation}
where $\hat{N} = \hat{n}_1 + \hat{n}_2$ is the total boson number
operator. The equation (\ref{h-su2-casimir}) shows clearly that
${\cal C}$ depends only on $\hat{N}$.

Calculating the expectation value $\langle n_1 n_2 | {\cal C} |
n_1 n_2 \rangle $ and comparing it with Eq. (\ref{urep}), we have
\begin{equation}
    \tilde{j} = \frac12 (N + 2 \alpha).
\end{equation}
The fact that the values of $\tilde{j}$ are half integers
($\tilde{j}=0$, $1/2$, 1, ...) requires $\alpha = 0$, thus, the
irreducible representation $\tilde{j}$ of ${\cal H}$ is characterized by the
total boson number $N$, namely, $\tilde{j} = N / 2$. The similar
conclusion exists for SU(2).\cite{bl} Correspondingly, Eq. (\ref{su2-11-1})
leads to the simplest form
\begin{eqnarray}
\begin{array}{rl}
\dot{B}^{(1,1)}_2 ({\cal J}_+) = &  \sqrt{ \frac12 C_1 +
    \frac18 C_3 [ \hat{n}_1^2 + \hat{n}_2 ( \hat{n}_2 + 2) ] } a_1^+ a_2,    \\
\dot{B}^{(1,1)}_2 ({\cal J}_-) = & a_1 a^+_2 \sqrt{ \frac12 C_1 +
    \frac18 C_3 [ \hat{n}_1^2 + \hat{n}_2 ( \hat{n}_2 + 2) ] },    \\
\dot{B}^{(1,1)}_2 ({\cal J}_3)  = & \frac{1}{2} (\hat{n}_1 - \hat{n}_2) ,
\end{array}
\label{su2-11-2}
\end{eqnarray}
which may also be obtained by considering the second solution (\ref{solu-2}) with setting
$\alpha = 0$.
When $C_1=2$ and $C_3=0$, Eq. (\ref{su2-11-2}) becomes the standard Jordan-Schwinger
realization (\ref{su2}) of SU(2).

Now discuss the properties of the spaces that $\dot{B}^{(1,1)}_2 ({\cal J}_{\mu})$
($\mu = \pm $, 3) act on. We observe that for $C_3 \not= 0$
the square-root symbols appear in the two-boson realization
(\ref{su2-11-2}), which is analogous to the Holstein-Primakoff
single-boson realization of SU(2). \cite{hp-1}
The acting spaces of $\dot{B}^{(1,1)}_2 ({\cal J}_{\mu})$ may be
certain subspaces of the Fock space
${\cal F} = \{\left| n_1 n_2 \right \rangle \:| \:n_1, n_2=0,1,2,... $\},
in which $n_1$ and $n_2$ need limiting in order that the values of
the square roots appeared in the matrix elements
$\langle n_1 \pm 1 n_2 \mp 1 | \dot{B}^{(1,1)}_2 ({\cal J}_{\pm}) | n_1 n_2 \rangle$
must be greater than or equal to zero. For the realization
(\ref{su2-11-2}), $n_1$ and $n_2$ have to satisfy the constraint
conditions
\begin{eqnarray}
   \left \{
    \begin{array}{l}
    (n_1 + 1)^2 + n_2^2 \geq 1 - 4 \frac{C_1}{C_3},  \\
    n_1^2 + (n_2 + 1)^2 \geq 1 - 4 \frac{C_1}{C_3}.
    \end{array}
  \right.
\label{cc-1}
\end{eqnarray}
The results of Eq. (\ref{cc-1}), which are pertinent to the relative signs of
$C_1$ and $C_3$, may be put into the following two categories.

(A) If $C_1$ has the same sign as $C_3$, then Eq. (\ref{cc-1})
always holds so that the acting space of $\dot{B}^{(1,1)}_2 ({\cal
J}_{\mu})$ is the whole Fock space ${\cal F}$. In ${\cal F}$, the
infinite-dimensional nullspaces of $\dot{B}^{(1,1)}_2 ({\cal
J}_+)$ and $\dot{B}^{(1,1)}_2 ({\cal J}_-)$ are
$$
    \left \{ | n_1 \, 0  \rangle \, | \, n_1=0, \, 1, ... \right \}
\hspace{3mm} \mbox{and} \hspace{3mm}
    \left \{ | 0 \, n_2  \rangle \, | \, n_2=0, \, 1, ... \right \},
$$
respectively, since they satisfy
$$
    \dot{B}^{(1,1)}_2 ({\cal J}_+) \, | n_1 \, 0  \rangle =
    \dot{B}^{(1,1)}_2 ({\cal J}_-) \, | 0 \, n_2  \rangle = 0.
$$
Obviously, $| 0  0 \rangle$ is the common nullspace state of
$\dot{B}^{(1,1)}_2 ({\cal J}_+)$ and $\dot{B}^{(1,1)}_2 ({\cal
J}_-)$

(B) If the sign of $C_1$ is opposite to that of $C_3$, then the values
of $n_1$ and $n_2$ are limited by Eq. (\ref{cc-1}).
Consider first that $n_1$ takes independently values, then the smallest value that
$n_2$ may take, which depends on $n_1$, should be
$\zeta_1 (n_1) \equiv \left [ \sqrt{ 1 - 4 C_1 / C_3 - ( n_1 +1)^2} \right ]$,
where the symbol $[x]$ for a real number $x$ means taking an integer greater than $x$,
so that the acting space of $\dot{B}^{(1,1)}_2 ({\cal J}_{\mu})$ is
$$
  \dot{V}_1 = \bigcup_{n_1 = 0}^{\eta} \dot{V}_1 (n_1) \subset {\cal F} ,
$$
where
$$
  \dot{V}_1 (n_1) \equiv
    \{ | n_1 , \zeta_1 (n_1)  +i \rangle \, | \, i=0, \, 1, ...\},
\hspace{5mm}
 \eta \equiv \left [ \sqrt{ 1 - 4 C_1 / C_3 } \right ] - 1.
$$
In $\dot{V}_1$, $\dot{V}_1 (0)$ is the infinite-dimensional
nullspace of $\dot{B}^{(1,1)}_2 ({\cal J}_-)$
since all the states in $\dot{V}_1 (0)$ satisfy $\dot{B}^{(1,1)}_2
({\cal J}_-) | 0 ,\zeta_1 (0) + i \rangle = 0$ ($i=0$, 1, ...).
The subspace \{$| n_1 , \zeta_1 (n_1) \rangle \: | \: n_1 = 0, 1,
..., \eta$\} in $\dot{V}_1$ is the ($\eta + 1$)-dimensional
nullspace of $\dot{B}^{(1,1)}_2 ({\cal J}_+)$, which satisfies
$\dot{B}^{(1,1)}_2 ({\cal J}_+) | n_1 , \zeta_1 (n_1) \rangle =
0$. Moreover, $| 0 , \zeta_1 (0) \rangle$ is the common nullspace
state of $\dot{B}^{(1,1)}_2 ({\cal J}_+)$ and $\dot{B}^{(1,1)}_2
({\cal J}_-)$.

In view of the simple symmetry $n_1 \leftrightarrow n_2$ between
the two equations of Eq. (\ref{cc-1}), if $n_2$ takes
independently values, then the smallest value of $n_1$ should be
$\zeta_2 (n_2) \equiv \left [ \sqrt{ 1 - 4 C_1 / C_3 - ( n_2
+1)^2} \right ]$, hence the acting space of $\dot{B}^{(1,1)}_2
({\cal J}_{\mu})$ is
$$
  \dot{V}_2 = \bigcup_{n_2 = 0}^{\eta} \dot{V}_2 (n_2)
    \equiv \bigcup_{n_2 = 0}^{\eta}
    \{ | \zeta_2 (n_2) +i , n_2 \rangle \, | \, i=0, \, 1, ...\}
  \subset {\cal F}.
$$
In $\dot{V}_2$, $\dot{V}_2 (0) $ is the infinite-dimensional
nullspace of $\dot{B}^{(1,1)}_2 ({\cal J}_+)$, \{$| \zeta_2 (n_2),
n_2 \rangle \: | \: n_2 = 0, 1, ..., \eta$\} is the ($\eta +
1$)-dimensional nullspace of $\dot{B}^{(1,1)}_2 ({\cal J}_-)$, and
$| \zeta_2 (0), 0 \rangle$ is the common nullspace state of
$\dot{B}^{(1,1)}_2 ({\cal J}_+)$ and $\dot{B}^{(1,1)}_2 ({\cal
J}_-)$.

(2) If the unitary relations need not satisfying, it follows from
Eq. (\ref{solu-1})
 that the conventional choice $\dot{g}^{(1,1)}_1 (\hat{n}_1,\hat{n}_2) = 1$
(or $\dot{f}^{(1,1)}_1 (\hat{n}_1,\hat{n}_2) = 1$) may immediately give rise to
a nonunitary two-boson realization
\begin{eqnarray}
\begin{array}{rl}
\dot{B}^{(1,1)}_3 ({\cal J}_+) = & \frac{1}{8 \hat{n}_1} (\hat{n}_1 + 2 \alpha )
         \{ 4 C_1 + C_3 [ \hat{n}_1 (\hat{n}_1 + 4 \alpha )  \\
    {}  &    + (\hat{n}_2 + 1)^2 + (2 \alpha +1)(2 \alpha -1)]  \}
    a_1^+ a_2,    \\
\dot{B}^{(1,1)}_3 ({\cal J}_-) = & a_1 a^+_2 ,    \\
\dot{B}^{(1,1)}_3 ({\cal J}_3)  = & \frac{1}{2} (\hat{n}_1 - \hat{n}_2) + \alpha.
\end{array}
\label{su2-11-3}
\end{eqnarray}
In terms of Eq. (\ref{su2-11-3}), the Casimir invariant ${\cal C}$, Eq. (\ref{casimir}),
of ${\cal H}$ has the same expression as Eq. (\ref{h-su2-casimir}). So taking $\alpha = 0$
in Eq. (\ref{su2-11-3}) leads to
\begin{eqnarray}
\begin{array}{rl}
\dot{B}^{(1,1)}_4 ({\cal J}_+) = & \left \{ \frac12 C_1 +
    \frac18 C_3 [ \hat{n}_1^2 + \hat{n}_2 ( \hat{n}_2 + 2 ) ] \right \}
    a_1^+ a_2,    \\
\dot{B}^{(1,1)}_4 ({\cal J}_-) = & a_1 a^+_2 ,    \\
\dot{B}^{(1,1)}_4 ({\cal J}_3)  = & \frac{1}{2} (\hat{n}_1 - \hat{n}_2).
\end{array}
\label{su2-11-4}
\end{eqnarray}
Different from the unitary realization (\ref{su2-11-2}),
no square-root symbols appear in the above nonunitary realization (\ref{su2-11-4}),
hence, it may not only avoid the convergence questions associated with the expansion
of square-root operator but also make the values of $n_1$ and $n_2$ in
\{$| n_1 n_2 \rangle$\} unlimited, i.e.,
the acting space of $\dot{B}^{(1,1)}_4 ({\cal J}_{\mu})$ is the whole Fock space.
Taking especially $C_1=2$ and $C_3=0$, Eq. (\ref{su2-11-4}) gives an unitary
realization of SU(2), i.e., the Jordan-Schwinger realization (\ref{su2}), while
taking $C_1=-2$ and $C_3=0$, Eq. (\ref{su2-11-4}) does a nonunitary realization of SU(1,1).
We notice that for $C_3 \not= 0$ the two-boson realization (\ref{su2-11-4}) is in fact
analogous to the Dyson single-boson realization of SU(2). \cite{dyson}

(3) Another nonunitary realization may be obtained by choosing
$\dot{g} (\hat{n}_1,\hat{n}_2) = \dot{f} (\hat{n}_1 - 1 ,\hat{n}_2 + 1) $
and $\alpha =0$ in Eq. (\ref{solu-1}) as
\begin{eqnarray}
\begin{array}{rl}
    \dot{B}^{(1,1)}_5 ({\cal J}_+) & = \dot{f} (\hat{n}_1,\hat{n}_2) a_1^+ a_2,  \\
    \dot{B}^{(1,1)}_5 ({\cal J}_-) & = \dot{f} (\hat{n}_1,\hat{n}_2) a_1 a_2^+ , \\
    \dot{B}^{(1,1)}_5 ({\cal J}_3) & = \frac{1}{2} (\hat{n}_1 - \hat{n}_2),
\end{array}
\label{su2-11-5}
\end{eqnarray}
where $\dot{f} (\hat{n}_1,\hat{n}_2)$ satisfies
\begin{equation}
   8 \dot{f} (\hat{n}_1,\hat{n}_2) =
        \left \{ 4 C_1 + C_3 [ n_1^2 + n_2 (n_2 + 2) ] \right \}
     \dot{f}^{-1} (\hat{n}_1 -1,\hat{n}_2+1).
\label{su2-f-eq}
\end{equation}
Note that here $ \dot{B}^{(1,1)}_5 ({\cal J}_{\pm}) \not=
( \dot{B}^{(1,1)}_5 ({\cal J}_{\mp}) )^{\dagger}$ for the real function
$\dot{f} (\hat{n}_1,\hat{n}_2)$.
We call Eq. (\ref{su2-11-5}) a constrained nonunitary realization since
$\dot{B}^{(1,1)}_5 ({\cal J}_+)$ and $\dot{B}^{(1,1)}_5 ({\cal J}_-)$
utilize the same function $\dot{f} (\hat{n}_1,\hat{n}_2)$.
With the help of Eq. (\ref{n12-eigen}), solving Eq. (\ref{su2-f-eq}) gives rise to
\begin{eqnarray}
\begin{array}{rl}
    \dot{f} (\hat{n}_1, \hat{n}_2) = &
    \exp \left \{ (-1)^{\hat{n}_1 - 1} \left [
        - \dot{\Omega}^{--}_1 (\hat{N})
        + \dot{\Omega}^{--}_3 (\hat{N})
        - \dot{\Omega}^{-+}_1 (\hat{N})
        + \dot{\Omega}^{-+}_3 (\hat{N})
    \right. \right. \\
    {} &    \left. \left.
        + (-1)^{\hat{n}_1} \left ( \dot{\Omega}^{+-}_1 (\hat{M})
        - \dot{\Omega}^{+-}_3 (\hat{M})
        + \dot{\Omega}^{++}_1 (\hat{M})
        - \dot{\Omega}^{++}_3 (\hat{M}) \right )
    \right. \right. \\
    {} &    \left. \left.
    + \frac12 \ln (C_3) [1 - (-1)^{\hat{n}_1} ]+ \dot{v}(\hat{N})
    \right ] \right \}
\end{array}
\label{su2-f-solu}
\end{eqnarray}
where $\dot{v}(\hat{N})$ is an arbitrary function of $\hat{N}$, and
\begin{equation}
 \dot{\Omega}^{\pm \pm}_k ( \hat{x}) \equiv \ln \left \{ \Gamma \left [
    \frac14 \left ( k \pm \hat{x} \pm
    \sqrt{ -8 C_1 / C_3 - (\hat{N}^2 + 2 \hat{N} -1)} \right ) \right ] \right \},
\label{omega}
\end{equation}
in which the order of two superscripts $\pm \pm$ of $\dot{\Omega}$
is the same as that of them appearing in the equation of r.h.s.,
and $\Gamma[a(\hat{N})]$ is an operator function, whose
expectation value in ${\cal F}$ in fact is the ordinary Gamma
function $\Gamma[a(N)]$ for the real number $a(N)$, i.e.,
\begin{equation}
    \langle n_1 n_2 |\, \Gamma[a(\hat{N})] \, | n_1 n_2 \rangle
     = \Gamma[a(N)].
\label{gamma}
\end{equation}
Different from the nonunitary realization (\ref{su2-11-4}),
this nonunitary realization (\ref{su2-11-5}) may not be reduced to
the Jordan-Schwinger realization (\ref{su2}) of SU(2) since
in Eqs. (\ref{su2-f-solu}) and (\ref{omega}) $C_3$ can not take zero.

It will be verified later that the nonunitary realizations (\ref{su2-11-4}) and
(\ref{su2-11-5}) may be connected with the unitary realization (\ref{su2-11-2}) by
similarity transformations.

\vspace{2mm}
{\bf 2. The $(2,2)$ case.}

Setting $k=l=2$ in Eq. (\ref{fg-su2}) and taking $\alpha = 0$ into
account, we may obtain two solutions. One of them is given by
\begin{eqnarray}
\begin{array}{rl}
    \dot{f}^{(2,2)}_1  \dot{g}^{(2,2)}_1 = &
     [ 128 ( \hat{n}_1 - X_{1}^+ (\hat{n}_1) /2) (\hat{n}_2 + 1 )(\hat{n}_2 + 2 )]^{-1} \\
    {}  & ( \hat{n}_2 + X_{3}^+ (\hat{n}_1) /2)
        \{ 16 C_1 + C_3 [ \hat{n}_1^2 - X_{1}^- (\hat{n}_1) (\hat{n}_1 + 1)  \\
    {}  &   + \hat{n}_2 ( \hat{n}_2 + X_{3}^+ (\hat{n}_1)) ] \},
\end{array}
\label{solu-3}
\end{eqnarray}
where
\begin{equation}
    X_{k}^{\pm}(\hat{n}_1) \equiv k \pm (-1)^{\hat{n}_1}.
\label{def-x}
\end{equation}
Another solution may be directly get from Eq. (\ref{solu-3}) by considering
the symmetry $ n_1 \leftrightarrow n_2$.
In the same way as discussing the $(1,1)$ case, in terms of Eq. (\ref{solu-3}),
we may obtain the unitary two-boson realization of quadratic type
\begin{eqnarray}
\begin{array}{rl}
\dot{B}^{(2,2)}_1 ({\cal J}_+) = &
    [ 128 ( \hat{n}_1 - X_{1}^+ (\hat{n}_1) /2) (\hat{n}_2 + 1 )(\hat{n}_2 + 2 )]^{-1/2} \\
    {}  & \{ ( \hat{n}_2 + X_{3}^+ (\hat{n}_1) /2)
        \{ 16 C_1 + C_3 [ \hat{n}_1^2 - X_{1}^- (\hat{n}_1) (\hat{n}_1 + 1)  \\
    {}  &   + \hat{n}_2 ( \hat{n}_2 + X_{3}^+ (\hat{n}_1)) ] \} \}^{1/2}
    (a_1^+)^2 a_2^2,    \\
\dot{B}^{(2,2)}_1 ({\cal J}_-) = & a_1^2 (a_2^+)^2
    [ 128 ( \hat{n}_1 - X_{1}^+ (\hat{n}_1) /2) (\hat{n}_2 + 1 )(\hat{n}_2 + 2 )]^{-1/2} \\
    {}  & \{ ( \hat{n}_2 + X_{3}^+ (\hat{n}_1) /2)
        \{ 16 C_1 + C_3 [ \hat{n}_1^2 - X_{1}^- (\hat{n}_1) (\hat{n}_1 + 1)  \\
    {}  &   + \hat{n}_2 ( \hat{n}_2 + X_{3}^+ (\hat{n}_1)) ] \} \}^{1/2} ,  \\
\dot{B}^{(2,2)}_1 ({\cal J}_3)  = & \frac{1}{4} (\hat{n}_1 - \hat{n}_2)
\end{array}
\label{su2-22-1}
\end{eqnarray}
and the nonunitary two-boson realization of quadratic type
\begin{eqnarray}
\begin{array}{rl}
\dot{B}^{(2,2)}_2 ({\cal J}_+) = &
    [ 128 ( \hat{n}_1 - X_{1}^+ (\hat{n}_1) /2) (\hat{n}_2 + 1 )(\hat{n}_2 + 2 )]^{-1} \\
    {}  & \{( \hat{n}_2 + X_{3}^+ (\hat{n}_1) /2)
        \{ 16 C_1 + C_3 [ \hat{n}_1^2 - X_{1}^- (\hat{n}_1) (\hat{n}_1 + 1)  \\
    {}  &   + \hat{n}_2 ( \hat{n}_2 + X_{3}^+ (\hat{n}_1)) ] \} \}^{1/2}
    (a_1^+)^2 a_2^2,    \\
\dot{B}^{(2,2)}_2 ({\cal J}_-) = & a_1^2 (a_2^+)^2 ,    \\
\dot{B}^{(2,2)}_2 ({\cal J}_3)  = & \frac{1}{4} (\hat{n}_1 - \hat{n}_2).
\end{array}
\label{su2-22-2}
\end{eqnarray}

We observe that the unitary realization (\ref{su2-22-1}) is
explicitly different from that of Ref. \cite{bbd}, in which
$J_{\pm}$ and $J_3$ are first defined as $J_+ = (a_1^+)^k a_2^l$,
$J_- = a_1^k (a_2^+)^l$ and $J_3 = (\hat{n}_1 - \hat{n}_2)/(k+l)$,
however, in order to generate ${\cal H}$ the unique
non-trivial choice is $k=l=2$, combined with the coefficient
of $J_3^3$, in the commutator $[J_+, \: J_-]$, being the fixed
number $-64$, and the coefficient of $J_3$ in fact being the operator
function of $\hat{N}$. However, the realization defined by Eq.
(\ref{h-su2-jj}) allows the arbitrary powers and constant coefficients.

\subsection{The second kind of realizations}

In analogy with the Jordan-Schwinger realization (\ref{su11}) of SU(1,1),
the second kind of two-boson realizations of ${\cal H}$ may be constructed
in the following scheme:
\begin{equation}
\begin{array}{rl}
    \ddot{B}^{(k,l)} ({\cal J}_+) & = \ddot{f} (\hat{n}_1,\hat{n}_2)
        (a_1^+)^k (a_2^+)^l,  \\
    \ddot{B}^{(k,l)} ({\cal J}_-) & = a_1^k a_2^l \, \ddot{g}
        (\hat{n}_1,\hat{n}_2), \\
    \ddot{B}^{(k,l)} ({\cal J}_3) & = \ddot{h} (\hat{n}_1,\hat{n}_2),
\end{array}
\label{h-su11-jj}
\end{equation}
where $k$ and $l$ are positive integers, the operator functions
$\ddot{f} (\hat{n}_1,\hat{n}_2)$, $\ddot{g} (\hat{n}_1,\hat{n}_2)$ and
$\ddot{h} (\hat{n}_1,\hat{n}_2)$ have to be determined
by the commutation relations (\ref{higgs}) of ${\cal H}$.
Acting $\ddot{B}^{(k,l)} ({\cal J}_{\pm})$ for a fixed $(k,l)$ on some basis vector
$| n_1 n_2 \rangle$ of ${\cal F}$ produces another basis vector
$| n_1 \pm k, n_2 \pm l\rangle$.

It follows that inserting Eq. (\ref{h-su11-jj}) into the first
equation of Eq. (\ref{higgs}) leads to the difference equation
\begin{equation}
    \ddot{h} (\hat{n}_1,\hat{n}_2) - \ddot{h} (\hat{n}_1 - k ,\hat{n}_2 - l) =1.
\label{h-equ-2}
\end{equation}
Its solution reads
\begin{equation}
    \ddot{h} (\hat{n}_1,\hat{n}_2) = \frac{\hat{n}_1}{2 k} + \frac{\hat{n}_2}{2 l} + \beta,
\label{h-solu-2}
\end{equation}
where the real constant $\beta$ will be determined later.

Using Eq. (\ref{h-solu-2}), in order to satisfy the second equation of Eq. (\ref{higgs}),
the following difference equation must hold:
\begin{equation}
\begin{array}{l}
    \left [ \prod\limits_{i=1}^{k} (\hat{n}_1 - i +1) \right ]
    \left [ \prod\limits_{i=1}^{l} (\hat{n}_2 - i +1) \right ]
     \ddot{f}^{(k,l)} (\hat{n}_1,\hat{n}_2) \ddot{g}^{(k,l)}(\hat{n}_1,\hat{n}_2)   \\
    - \left [ \prod\limits_{i=1}^{k} (\hat{n}_1 + i) \right ]
    \left [ \prod\limits_{i=1}^{l} (\hat{n}_2 + i) \right ]
     \ddot{f}^{(k,l)}(\hat{n}_1 + k ,\hat{n}_2 + l)
    \ddot{g}^{(k,l)}(\hat{n}_1 + k ,\hat{n}_2 + l) \\
    = C_1 \left ( \frac{\hat{n}_1}{2 k} + \frac{\hat{n}_2}{2 l} + \beta \right )
    + C_3  \left ( \frac{\hat{n}_1}{2 k} + \frac{\hat{n}_2}{2 l} + \beta \right )^3.
\label{fg-su11}
\end{array}
\end{equation}

Just like the first kind of realizations discussed in the last
subsection, in what follows, we will study the case of $(k,l)=(1,1)$,
and give directly the results of $(k,l)=(2,2)$.

\vspace{2mm}
{\bf 1. The $(1,1)$ case.}

Solving Eq. (\ref{fg-su11}) with setting $k=l=1$, we have two solutions:
\begin{eqnarray}
\begin{array}{rl}
    \ddot{f}^{(1,1)}_1 (\hat{n}_1,\hat{n}_2) \ddot{g}^{(1,1)}_1 (\hat{n}_1,\hat{n}_2) = &
        - \frac{1}{8 \hat{n}_1} (\hat{n}_1 + 2 \beta -1)
          \{ 4 C_1 + C_3 [ \hat{n}_1 (\hat{n}_1 + 4 \beta -2)  \\
    {}  &   + \hat{n}_2^2 + 4 \beta ( \beta -1)] \},
\end{array}
\label{su11-solu-1}
\end{eqnarray}
and
\begin{eqnarray}
\begin{array}{rl}
    \ddot{f}^{(1,1)}_2 (\hat{n}_1,\hat{n}_2) \ddot{g}^{(1,1)}_2 (\hat{n}_1,\hat{n}_2) = &
        - \frac{1}{8 \hat{n}_2} (\hat{n}_2 + 2 \beta -1)
          \{ 4 C_1 + C_3 [ \hat{n}_1^2 + \hat{n}_2 (\hat{n}_2 + 4 \beta -2)  \\
    {}  &   + 4 \beta ( \beta -1)] \} .
\end{array}
\label{su11-solu-2}
\end{eqnarray}
Between the two solutions there exists explicitly the symmetry:
$\hat{n}_1 \leftrightarrow \hat{n}_2$, so we need merely to consider
the solution (\ref{su11-solu-1}).

(1) If the unitary relations
$ \ddot{B}^{(1,1)} ({\cal J}_{\pm}) = (\ddot{B}^{(1,1)} ({\cal J}_{\mp}))^{\dagger}$
are imposed, namely,
$\ddot{f}^{(1,1)}_1 (\hat{n}_1,\hat{n}_2) = \ddot{g}^{(1,1)}_1 (\hat{n}_1,\hat{n}_2)$,
then solving Eq. (\ref{su11-solu-1}) and substituting it into Eq. (\ref{h-su11-jj}),
we obtain
\begin{eqnarray}
\begin{array}{rl}
\ddot{B}^{(1,1)}_1 ({\cal J}_+) = & \{ - \frac{1}{8 \hat{n}_1} (\hat{n}_1 + 2 \beta -1)
          \{ 4 C_1 + C_3 [ \hat{n}_1 (\hat{n}_1 + 4 \beta -2)  \\
    {}  &   + \hat{n}_2^2 + 4 \beta ( \beta -1)] \} \}^{1/2}
        a_1^+ a_2^+,    \\
\ddot{B}^{(1,1)}_1 ({\cal J}_-) = & a_1 a_2 \{
    - \frac{1}{8 \hat{n}_1} (\hat{n}_1 + 2 \beta -1)
          \{ 4 C_1 + C_3 [ \hat{n}_1 (\hat{n}_1 + 4 \beta -2)  \\
    {}  &   + \hat{n}_2^2 + 4 \beta ( \beta -1)] \} \}^{1/2},    \\
\ddot{B}^{(1,1)}_1 ({\cal J}_3)  = & \frac{1}{2} (\hat{n}_1 + \hat{n}_2) + \beta.
\end{array}
\label{su11-11-1}
\end{eqnarray}

Substituting Eq. (\ref{su11-11-1}) into Eq. (\ref{casimir}), the
Casimir invariant ${\cal C}$ of ${\cal H}$ may be expressed in
terms of $n_1$ and $n_2$ as
\begin{equation}
    {\cal C} = \frac{1}{64} (\hat{M} + 2 \beta -2 ) (\hat{M} + 2 \beta )
        [8 C_1 + C_3 (\hat{M} + 2 \beta -2 ) (\hat{M} + 2 \beta ) ],
\label{h-su11-casimir}
\end{equation}
where $\hat{M} = \hat{n}_1 - \hat{n}_2 $ or $\hat{n}_2 - \hat{n}_1 $ is
the number {\it difference} operator for two kinds of different bosons, while in
Eq. (\ref{h-su2-casimir}), the boson number {\it sum} operator, i.e.,
the total boson number operator $\hat{N}$, appears.
Calculating $\langle n_1 n_2 | {\cal C} | n_1 n_2 \rangle$ and then comparing it
with the forth equation of Eq. (\ref{urep}) gives
\begin{equation}
    \tilde{j} = \frac12 (M + 2 \beta - 2 ) \hspace{3mm} \mbox{or} \hspace{3mm}
    \tilde{j} = \frac12 (M - 2 \beta ) ,
\end{equation}
where $M = n_1 - n_2 $ or $n_2 - n_1 $ is the eigenvalue of $\hat{M}$.
The symmetry requires that $\beta= 1/2$, thus, the irreducible representation
$\tilde{j}$ of ${\cal H}$ are related to $M$ through the equation
$\tilde{j} = \frac12 (M - 1)$.
SU(1,1) has the similar result. \cite{bl}
Correspondingly, Eq. (\ref{su11-11-1}) becomes
\begin{eqnarray}
\begin{array}{rl}
\ddot{B}^{(1,1)}_2 ({\cal J}_+) = & \sqrt{ - \frac12 C_1 - \frac18 C_3 ( \hat{n}_1^2
    + \hat{n}_2^2 -1 ) } a_1^+ a_2^+,    \\
\ddot{B}^{(1,1)}_2 ({\cal J}_-) = & a_1 a_2
    \sqrt{ - \frac12 C_1 - \frac18 C_3 ( \hat{n}_1^2
    + \hat{n}_2^2 -1 ) },   \\
\ddot{B}^{(1,1)}_2 ({\cal J}_3)  = & \frac{1}{2} (\hat{n}_1 + \hat{n}_2 +1 ) .
\end{array}
\label{su11-11-2}
\end{eqnarray}
Thus, for $C_3 \not= 0$ the spaces that the operators
$\ddot{B}^{(1,1)}_2 ({\cal J}_{\mu})$ ($\mu = \pm $, 3) act on may
be certain subspaces of the Fock space ${\cal F} = \{\left| n_1
n_2 \right \rangle \:| \:n_1, n_2=0,1,2,... $\}, that is, $n_1$
and $n_2$ need limiting in order that the values of the square
roots appeared in the matrix elements $\langle n_1 \pm 1 n_2 \pm 1
| \ddot{B}^{(1,1)}_2 ({\cal J}_{\pm}) | n_1 n_2 \rangle$ must be
greater than or equal to zero. For the realization
(\ref{su11-11-2}), $n_1$ and $n_2$ have to satisfy the constraint
equation
\begin{equation}
    n_1^2 + n_2^2 \geq 1 -4 \frac{C_1}{C_3},
\label{cc-2}
\end{equation}
whose results are listed as follows.

(A) If $C_1 \geq C_3 / 4$, then Eq. (\ref{cc-2}) always holds, so that
the acting space of $\ddot{B}^{(1,1)}_2 ({\cal J}_{\mu})$
is the whole Fock space ${\cal F}$, in which
$$
\{|0 \, n_2 \rangle \, | \, n_2 =0,\, 1, \, 2,...\}
\hspace{3mm} \mbox{and} \hspace{3mm}
\{|n_1 \, 0 \rangle \, | \, n_1=0,\, 1, \, 2,...\}
$$
are the infinite-dimensional nullspaces of $\ddot{B}^{(1,1)}_2 ({\cal J}_-)$,
since they satisfy
$$
    \ddot{B}^{(1,1)}_2 ({\cal J}_-) |0 \, n_2 \rangle =
    \ddot{B}^{(1,1)}_2 ({\cal J}_-) |n_1 \, 0 \rangle =0.
$$

(B) If $C_1 < C_3 / 4$, then the values of $n_1$ and $n_2$ need limiting.
First consider that $n_1$ takes independently values, then the values that $n_2$
may take are dependent on $n_1$, especially, its smallest values should be
$\kappa_1 (n_1) \equiv \left [ \sqrt{ 1 - 4 C_1 / C_3 - n_1^2} \right ]$ for
the given $n_1$.
As a result, the acting subspace of $\dot{B}^{(1,1)}_2 ({\cal J}_{\mu})$ is
$$
  \ddot{V}_1 = \bigcup_{n_1 = 0}^{\lambda} \ddot{V}_1 (n_1),
$$
where
$$
  \ddot{V}_1 (n_1) \equiv
    \{ | n_1 , \kappa_1 (n_1) + i \rangle \, | \, i=0, \, 1, ...\},
\hspace{5mm}
  \lambda \equiv \left [ \sqrt{ 1 - 4 C_1 / C_3 } \right ] - 1.
$$
In $\ddot{V}_1$, $\ddot{B}^{(1,1)}_2 ({\cal J}_-)$ has an infinite-dimensional nullspace
$\ddot{V}_1(0)$ and a $\lambda$-dimensional nullspace
\{$| n_1 , \kappa_1 (n_1) \rangle \, | \, n_1= 1, 2, ..., \lambda$\}.

Secondly, $n_2$ takes independently values, by means of the symmetry
$n_1 \leftrightarrow n_2 $ of Eq. (\ref{cc-2}), then the smallest value of $n_1$
is $\kappa_2 (n_2) \equiv \left [ \sqrt{ 1 - 4 C_1 / C_3 - n_2^2} \right ]$,
so that the acting space of $\dot{B}^{(1,1)}_2 ({\cal J}_{\mu})$ is
$$
  \ddot{V}_2 = \bigcup_{n_2 = 0}^{\lambda} \ddot{V}_2 (n_2)
    \equiv \bigcup_{n_2 = 0}^{\lambda}
    \{ | \kappa_2(n_2) +i, n_2 \rangle \, | \, i=0, \, 1, ...\}.
$$
Obviously, in $\ddot{V}_2$, $\ddot{V}_2(0)$ and
\{$|\kappa_2(n_2), n_2 \rangle \, | \, n_2= 1, 2, ..., \lambda$\} are
the nullspaces of $\ddot{B}^{(1,1)}_2 ({\cal J}_-)$
with infinite-dimension and $\lambda$-dimension, respectively.

However, for the second kind of realization (\ref{su11-11-2}),
$\ddot{B}^{(1,1)}_2 ({\cal J}_+)$ and $\ddot{B}^{(1,1)}_2 ({\cal J}_-)$
have no the common nullspace state.

(2) If the unitary relations need not satisfying, it follows from Eq. (\ref{su11-11-1})
that the conventional choice $\ddot{g}^{(1,1)}(\hat{n}_1,\hat{n}_2) = 1$
(or $\ddot{f}^{(1,1)}(\hat{n}_1,\hat{n}_2) = 1$) results in
the following nonunitary two-boson realization
\begin{equation}
\begin{array}{rl}
\ddot{B}^{(1,1)}_3 ({\cal J}_+) = &  - \frac{1}{8 \hat{n}_1} (\hat{n}_1 + 2 \beta -1)
          \{ 4 C_1 + C_3 [ \hat{n}_1 (\hat{n}_1 + 4 \beta -2)  \\
    {}  &   + \hat{n}_2^2 + 4 \beta ( \beta -1)] \}
        a_1^+ a_2^+,    \\
\ddot{B}^{(1,1)}_3 ({\cal J}_-) = & a_1 a_2 ,    \\
\ddot{B}^{(1,1)}_3 ({\cal J}_3)  = & \frac{1}{2} (\hat{n}_1 + \hat{n}_2) + \beta.
\end{array}
\label{su11-11-3}
\end{equation}
Taking $\beta = 1/2 $, Eq. (\ref{su11-11-3}) becomes
\begin{equation}
\begin{array}{rl}
\ddot{B}^{(1,1)}_4 ({\cal J}_+) = &  - \left[ \frac12 C_1 + \frac18 C_3 ( \hat{n}_1^2 +
    \hat{n}_2^2 -1 ) \right ]   a_1^+ a_2^+,    \\
\ddot{B}^{(1,1)}_4 ({\cal J}_-) = & a_1 a_2 ,    \\
\ddot{B}^{(1,1)}_4 ({\cal J}_3)  = & \frac{1}{2} (\hat{n}_1 + \hat{n}_2 + 1).
\end{array}
\label{su11-11-4}
\end{equation}
When $C_1=-2$ and $C_3=0$, Eq. (\ref{su11-11-4}), together with Eq. (\ref{su11-11-2}),
recovers the unitary Jordan-Schwinger realization (\ref{su11}) of SU(1,1).

(3) Choosing $\ddot{g} (\hat{n}_1,\hat{n}_2) = \ddot{f} (\hat{n}_1 - 1 ,\hat{n}_2 - 1)$
and $\beta = 1/2$ in Eq. (\ref{su11-solu-2}), we may obtain another constrained nonunitary
realization
\begin{eqnarray}
\begin{array}{rl}
    \ddot{B}^{(1,1)}_5 ({\cal J}_+) & = \ddot{f} (\hat{n}_1,\hat{n}_2) a_1^+ a_2^+,  \\
    \ddot{B}^{(1,1)}_5 ({\cal J}_-) & = \ddot{f} (\hat{n}_1,\hat{n}_2) a_1 a_2 , \\
    \ddot{B}^{(1,1)}_5 ({\cal J}_3) & = \frac{1}{2} (\hat{n}_1 + \hat{n}_2 +1),
\end{array}
\label{su11-11-5}
\end{eqnarray}
where $\ddot{f} (\hat{n}_1,\hat{n}_2)$ obeys
\begin{equation}
   8 \ddot{f} (\hat{n}_1,\hat{n}_2) = - \left [ 4 C_1 +  C_3 ( n_1^2 + n_2^2 -1 ) \right ]
     \ddot{f}^{-1} (\hat{n}_1 -1,\hat{n}_2 -1),
\label{su11-f-eq}
\end{equation}
whose solution is
\begin{eqnarray}
\begin{array}{rl}
    \ddot{f} (\hat{n}_1, \hat{n}_2) = &
    \exp \left \{ (-1)^{\hat{n}_1 - 1} \left [
        - \ddot{\Omega}^{--}_2 (\hat{M})
        + \ddot{\Omega}^{--}_4 (\hat{M})
        - \ddot{\Omega}^{-+}_2 (\hat{M})
        + \ddot{\Omega}^{-+}_4 (\hat{M})
    \right. \right. \\
    {} &    \left. \left.
        + (-1)^{\hat{n}_1} \left (
        \ddot{\Omega}^{+-}_2 (\hat{N})
        - \ddot{\Omega}^{+-}_4 (\hat{N})
        + \ddot{\Omega}^{++}_2 (\hat{N})
        - \ddot{\Omega}^{++}_4 (\hat{N}) \right )
    \right. \right. \\
    {} &    \left. \left.
    + \frac12 ( \ln (C_3) + \mbox{i} \pi )[1 - (-1)^{\hat{n}_1} ]
    + \ddot{v}(\hat{M})
    \right ] \right \},
\end{array}
\label{su11-f-solu}
\end{eqnarray}
where $\ddot{v}(\hat{M})$ is an arbitrary function of $\hat{M}$, and
\begin{equation}
 \ddot{\Omega}^{\pm \pm}_k ( \hat{x}) \equiv \ln \left \{ \Gamma \left [
    \frac14 \left ( k \pm \hat{x} \pm
    \sqrt{ -8 C_1 / C_3 - (\hat{M}^2 - 2)} \right ) \right ] \right \},
\label{omega2}
\end{equation}
in which $\Gamma[a(\hat{N})]$ has been defined by Eq. (\ref{gamma}).
The nonunitary realization (\ref{su11-11-5}) can not become the
Jordan-Schwinger realization (\ref{su11}) of SU(1,1) on
account of the singularity of $C_3$ in Eqs. (\ref{su11-f-solu})
and (\ref{omega2}).

We notice that all the nonunitary realizations, (\ref{su2-11-4}), (\ref{su2-11-5}),
(\ref{su11-11-4}) and (\ref{su11-11-5}), obtained above
are different from the inhomogeneous two-boson realizations obtained
in Ref. \cite{ruan1} by using the boson mapping method based upon
the induced representations of ${\cal H}$ on the quotient spaces
$U({\cal H})/I_i$ ($i=1$, 2), where $U({\cal H})$ is
the universal enveloping algebra of ${\cal H}$ and
$I_i$ are two left ideals with respect to $U({\cal H})$.

\vspace{2mm}
{\bf 2. The $(2,2)$ case.}

Equation (\ref{fg-su11}) with setting $k=l=2$ and $\beta=1/2$ has two solutions,
the first one is given by
\begin{eqnarray}
\begin{array}{rl}
    \ddot{f}^{(2,2)}_1  \ddot{g}^{(2,2)}_1 = &
         [ 128 ( \hat{n}_1 - 1) \hat{n}_1 (\hat{n}_2 -1 ) n_2 ]^{-1}
        ( \hat{n}_1 + X_{1}^- (\hat{n}_1) /2)   \\
    {}  &   ( \hat{n}_2 + 2 - X_{5}^- (\hat{n}_1) /2)
        \{ 16 C_1 + C_3 [ X_{1}^- (\hat{n}_1) (\hat{n}_1 +3) +
        \hat{n}_1^2  \\
    {}  &   + \hat{n}_2 (\hat{n}_2 + 4 + X_{3}^+ (\hat{n}_1))
        + 2 (2 - X_{3}^+ (\hat{n}_1) ) ] \},
\end{array}
\label{su11-fg3}
\end{eqnarray}
where the symbol $X_{k}^{\pm}(\hat{n}_1)$ has be defined by
Eq. (\ref{def-x}). The second solution may be directly obtained from
Eq. (\ref{su11-fg3}) by the substitutions $n_1 \rightarrow n_2$ and
$n_2 \rightarrow n_1$.
Solving Eq. (\ref{su11-fg3}) by considering respectively the unitary and
nonunitary conditions, and then inserting them into Eq. (\ref{h-su11-jj}),
we may obtain for ${\cal H}$ the unitary two-boson realization of quadratic
type
\begin{eqnarray}
\begin{array}{rl}
\ddot{B}^{(2,2)}_1 ({\cal J}_+) = &
    [ 128 ( \hat{n}_1 - 1) \hat{n}_1 (\hat{n}_2 -1 ) n_2 ]^{-1/2}
    [ \hat{n}_1 + X_{1}^- (\hat{n}_1) /2 ]   \\
    {}  &   [ \hat{n}_2 + 2 - X_{5}^- (\hat{n}_1) /2 ]
        \{ 16 C_1 + C_3 [ X_{1}^- (\hat{n}_1) (\hat{n}_1 +3) +
        \hat{n}_1^2  \\
    {}  &   + \hat{n}_2 (\hat{n}_2 + 4 + X_{3}^+ (\hat{n}_1))
        + 2 (2 - X_{3}^+ (\hat{n}_1) ) ] \}^{1/2}
    (a_1^+)^2 (a_2^+)^2,    \\
\dot{B}^{(2,2)}_1 ({\cal J}_-) = & a_1^2 a_2^2
    [ 128 ( \hat{n}_1 - 1) \hat{n}_1 (\hat{n}_2 -1 ) n_2 ]^{-1/2}
    [ \hat{n}_1 + X_{1}^- (\hat{n}_1) /2 ]   \\
    {}  &  [ \hat{n}_2 + 2 - X_{5}^- (\hat{n}_1) /2 ]
        \{ 16 C_1 + C_3 [ X_{1}^- (\hat{n}_1) (\hat{n}_1 +3) +
        \hat{n}_1^2  \\
    {}  &   + \hat{n}_2 (\hat{n}_2 + 4 + X_{3}^+ (\hat{n}_1))
        + 2 (2 - X_{3}^+ (\hat{n}_1) ) ] \}^{1/2}  ,    \\
\dot{B}^{(2,2)}_1 ({\cal J}_3)  = & \frac{1}{4} (\hat{n}_1 + \hat{n}_2 + 2)
\end{array}
\label{su11-22-1}
\end{eqnarray}
and the nonunitary two-boson realization of quadratic type
\begin{eqnarray}
\begin{array}{rl}
\ddot{B}^{(2,2)}_2 ({\cal J}_+) = &
    [ 128 ( \hat{n}_1 - 1) \hat{n}_1 (\hat{n}_2 -1 ) n_2 ]^{-1}
     [ \hat{n}_1 + X_{1}^- (\hat{n}_1) /2 ]   \\
    {}  &  [ \hat{n}_2 + 2 - X_{5}^- (\hat{n}_1) /2 ]
        \{ 16 C_1 + C_3 [ X_{1}^- (\hat{n}_1) (\hat{n}_1 +3) +
        \hat{n}_1^2  \\
    {}  &   + \hat{n}_2 (\hat{n}_2 + 4 + X_{3}^+ (\hat{n}_1))
        + 2 (2 - X_{3}^+ (\hat{n}_1) ) ] \}
    (a_1^+)^2 (a_2^+)^2,    \\
\dot{B}^{(2,2)}_2 ({\cal J}_-) = & a_1^2 a_2^2 ,    \\
\dot{B}^{(2,2)}_2 ({\cal J}_3)  = & \frac{1}{4} (\hat{n}_1 + \hat{n}_2 +2 ).
\end{array}
\label{su11-22-2}
\end{eqnarray}

\section{UNITARIZATION EQUATIONS AND SIMILARITY TRANSFORMATIONS}
\label{4}

In the last section, the two kinds of two-boson realizations of
${\cal H}$ are constructed, and in each kind, one unitary
realization and two different nonunitary realizations are
discussed, respectively. In this section, we will show that the
unitary realizations and the nonunitary realizations in the same
kind may be connected by similarity transformations.

Let us begin with discussing the general procedure. Denote the
unitary boson realization and the nonunitary boson realization by
$B^{\mbox{u}}({\cal J}_{\mu})$ ($\mu = \pm$, 3) and
$B^{\mbox{nu}}({\cal J}_{\mu})$, respectively, and the
corresponding similarity transformation by $S$, then we have
\begin{equation}
    S B^{\mbox{nu}} ({\cal J}_{\mu}) S^{-1} = B^{\mbox{u}} ({\cal J}_{\mu}).
\label{s-t}
\end{equation}
Hence, $S$ in general is an operator function with respect to
the boson operators and the particle number operators.

Using Eq. (\ref{s-t}) and the unitary conditions satisfied by $B^{\mbox{u}} ({\cal J}_{\mu})$
\begin{eqnarray}
\begin{array}{c}
   \left( B^{\mbox{u}} ({\cal J}_\pm) \right)^{\dagger} = B^{\mbox{u}}({\cal J}_\mp), \\
   \left( B^{\mbox{u}} ({\cal J}_3) \right)^{\dagger} = B^{\mbox{u}}({\cal J}_3),
\end{array}
\label{u-r}
\end{eqnarray}
it follows that we may obtain the following unitarization equations obeyed by
$B^{\mbox{nu}} ({\cal J}_{\mu})$
\begin{eqnarray}
\begin{array}{c}
    U^{-1} \left(B^{\mbox{nu}} ({\cal J}_{\pm}) \right)^{\dagger} U =
        B^{\mbox{nu}} ({\cal J}_{\mp}),  \\
    U^{-1} \left(B^{\mbox{nu}} ({\cal J}_3) \right)^{\dagger} U =
        B^{\mbox{nu}} ({\cal J}_3),
\end{array}
\label{u-t}
\end{eqnarray}
where $ U \equiv S^{\dagger} S $ is an Hermitian operator.
The similarity transformation $S$ may be obtained by solving Eq. (\ref{u-t}) in the
Fock space.

We observe from the two-boson realizations (\ref{su2-11-4}), (\ref{su2-11-5}),
(\ref{su11-11-4}) and (\ref{su11-11-5}) obtained in the last section that
$\dot{B}^{(1,1)}_4 ({\cal J}_3)$,
$\dot{B}^{(1,1)}_5 ({\cal J}_3)$,
$\ddot{B}^{(1,1)}_4 ({\cal J}_3)$
and $\ddot{B}^{(1,1)}_5 ({\cal J}_3)$
in fact are already Hermitian, so Eq. (\ref{s-t})
implies that the corresponding similarity transformations commute with ${\cal J}_3$,
in other words, they depend only on the particle number operators,
$\hat{n}_1$ and $\hat{n}_2$.

Now let us seek the similarity transformations $S_1$ and $S_2$
that correspond to the nonunitary realizations (\ref{su2-11-4}) and (\ref{su11-11-4}),
respectively.
Calculating the matrix elements of the unitarization equations
(see Eq. (\ref{u-t})) satisfied respectively by
$\dot{B}^{(1,1)}_4 ({\cal J}_-)$ and $\ddot{B}^{(1,1)}_4 ({\cal J}_-)$
in the Fock space ${\cal F}$,
and using Eqs. (\ref{su2-11-4}) and (\ref{su11-11-4}),
we may deduce the recurrent equations satisfied by the expectation values
$ S_i (n_1, n_2) \equiv \langle n_1 n_2 | S_i | n_1 n_2 \rangle$
($i=1$, 2),
\begin{equation}
    \left \{ 4 C_1 + C_3 [n_1^2 + n_2 (n_2 +2)] \right \} S_1 (n_1 , n_2) ^2 =
    8 S_1 (n_1-1, n_2+1) ^2,
\label{s1-eq}
\end{equation}
and
\begin{equation}
    \left [ 4 C_1 + C_3 ( n_1^2 + n_2^2 -1 ) \right ] S_2 (n_1 , n_2) ^2 =
    -8 S_1 (n_1 -1, n_2 -1) ^2.
\label{s2-eq}
\end{equation}
Solving Eqs. (\ref{s1-eq}) and (\ref{s2-eq}), and then using
Eq. (\ref{n12-eigen}), we obtain
\begin{equation}
    S_1 (\hat{n}_1, \hat{n}_2) = \sqrt{
    \frac{ ( C_3 / 4 )^{1 - \hat{n}_1} \dot{w}(\hat{N}) }
        {  ( \dot{Z} (\hat{N})_+ )_{\hat{n}_1 - 1}
        ( \dot{Z} (\hat{N})_+ )_{\hat{n}_1 - 1} }},
\label{s1}
\end{equation}
and
\begin{equation}
    S_2 (\hat{n}_1,\hat{n}_2) = \sqrt{ -
    \frac{ (-1)^{\hat{n}_1} ( C_3 /4 )^{1 - \hat{n}_1} \ddot{w}(\hat{M}) }
        { ( \ddot{Z} (\hat{M})_+ )_{\hat{n}_1 - 1}
        ( \ddot{Z} (\hat{M})_+ )_{\hat{n}_1 - 1} }},
\label{s2}
\end{equation}
respectively. In the above two equations, the minus signs out of the square-root symbols
have been omitted without loss of general,
$\dot{w}(\hat{N})$ and $\ddot{w}(\hat{M})$ are arbitrary functions with
respect to the sum operator $\hat{N}$ and the difference operator $\hat{M}$,
respectively,
\begin{equation}
   \dot{Z} (\hat{N})_{\pm} \equiv \frac12 \left [ 3- \hat{N} \pm
    \sqrt{ - 8 C_1 / C_3 - (\hat{N}^2 + 2 \hat{N} -1 ) } \right ],
\end{equation}
\begin{equation}
   \ddot{Z} (\hat{M})_{\pm} \equiv \frac12 \left [ 4- \hat{M} \pm
    \sqrt{- 8 C_1 / C_3 - (\hat{M}^2 - 2 ) } \right ],
\end{equation}
and the symbol $ (Z (\hat{N}))_{\hat{n}}$ in Eqs. (\ref{s1}) and (\ref{s2}) stands for
an operator function of $\hat{N}$, whose expectation value in ${\cal F}$ is the usual
Pochhammer symbol $(Z(N))_{n}$ for the real number $Z(N)$ and the positive integer $n$, i.e.,
\begin{equation}
    \langle n_1 n_2 | \, ( Z (\hat{N}) )_{\hat{n}} \, | n_1 n_2 \rangle
     = Z(N) [Z(N) +1] ... [Z(N) + n -1 ] \equiv (Z(N))_n.
\end{equation}

For the constrained nonunitary realizations (\ref{su2-11-5}) and
(\ref{su11-11-5}), there must exist the corresponding similarity transformations
$\bar{S}_1$ and $\bar{S}_2$, which connect Eq. (\ref{su2-11-5}) with
Eq. (\ref{su2-11-2}), and Eq. (\ref{su11-11-5}) with Eq. (\ref{su11-11-2}),
respectively. Using the same calculating method, we may obtain
\begin{equation}
    \bar{S}_1 (\hat{n}_1, \hat{n}_2) =  \sqrt{ \frac{8}{4 C_1 +
     C_3 [ \hat{n}_1^2 + \hat{n}_2 ( \hat{n}_2 + 2)]} },
\end{equation}
and
\begin{equation}
    \bar{S}_2 (\hat{n}_1, \hat{n}_2) = \sqrt{- \frac{8}{4 C_1 +
     C_3 ( \hat{n}_1^2 + \hat{n}_2^2 -1 ) } }.
\end{equation}

\section{SOME APPLICATIONS}
\label{5}
In this section, as applications, we shall apply the
results obtained previously to discussing the dynamical symmetry
of the Kepler system in the two-dimensional curved space and to
constructing phase operators of ${\cal H}$.

\subsection{Dynamical symmetry of the Kepler system in the two-dimensional curved space}
The key idea of dynamical symmetry is that the Hamiltonian
describing some quantum system can be constructed in terms of the
Casimir invariants, $C(g_1)$, $C(g_2)$, ..., of a chain of
algebras $g_1 \supset g_2 \supset ....$. \cite{iachello} The most
famous example of the dynamical symmetry is the nonrelativistic
hydrogen atom, \cite{pauli, fock, bargmann} whose Hamiltonian $H^{\mbox{c}}$
can be expressed by the first quadratic Casimir invariant, $C(\mbox{SO}(4))$,
of the SO(4) algebra, which is spanned by the three components of the angular momentum
${\bf J}$ and the three components of the Runge-Lentz-Laplace vector ${\bf R}$,
as $H^{\mbox{c}} \sim [ C(\mbox{SO}(4)) +1 ]^{-1}$. As mentioned in Sec. \ref{1},
Higgs has showed that the Kepler system in the two-dimensional
curved space is governed by the Higgs algebra ${\cal H}$, and
however, he applied the SO(3) algebra to calculate its energy
levels. In this subsection, we will show that the Hamiltonian
$H$ of this Kepler system may be naturally related to
the Casimir invariant ${\cal C}$ of ${\cal H}$, and then obtain
directly the energy levels of $H$ by using the
eigenvalue of ${\cal C}$.

The Hamiltonian of the Kepler system in the two-dimensional curved space
has the following expression \cite{higgs}
\begin{equation}
    H =  \frac12 \left ( \pi_i \pi_i + \lambda {\cal J}_3^2 \right) - \frac{\mu}{r},
\label{h-hami}
\end{equation}
where $\lambda$ is the curvature of the sphere, $\mu$ is a
constant number, ${\cal J}_3$ is a two-dimensional rotation
operator, and $\pi _i$ ($i=1$, 2), the two components of the momentum
operator $\vec{\pi}$ in the two-dimensional curved space, are
defined by
\begin{equation}
    \pi_i =  p_i - \frac{\lambda}{2}
        \left \{ x_i, \; ( {\bf x} \cdot {\bf p}) \right \},
\label{amo-cs}
\end{equation}
where \{ , \} is the usual anticommutator, $p_i= - \partial_{x_i}$
($i=1$, 2) are the two components of the ordinary momentum
operator ${\bf p}$ conjugate to ${\bf x}$, respectively.

This system possess three constants of motion: one is ${\cal J}_3$,
the remaining two are the two components of the Runge-Lentz-Laplace
vector ${\bf R}$ in the two-dimensional curved space, which, in
analogy with those in the three-dimensional flat space, \cite{pauli}
may be constructed as
\begin{equation}
    R_i = \frac12 \{{\cal J}_3, \; \epsilon_{ij} \pi_j  \}
    + \mu \frac{x_i}{r},  \hspace{4mm}  i=1, \:2,
\label{rl-cs}
\end{equation}
where $\epsilon_{ij}$ is the two-dimensional Levi-Civita symbol.

It can be easily verified that ${\cal J}_3$ and $R_{\pm} = R_1 \pm
\mbox{i} R_2$ satisfy
\begin{eqnarray}
\begin{array}{rl}
    [ {\cal J}_3, \: R_{\pm} ] = & \pm R_{\pm} ,  \cr
    [ R_+ , \: R_- ] = &  \left ( \frac{\lambda}{2}
        - 4 H \right ) {\cal J}_3
        + 4 \lambda {\cal J}_3^3,
\end{array}
\label{lr-cl}
\end{eqnarray}
and
\begin{equation}
    \{ R_+ , \: R_- \} =  2 \mu^2 + \left ( 2 H -
        \lambda {\cal J}_3^2 \right ) \left ( 2 {\cal J}_3^2 + \frac12 \right )
        - 2 \lambda {\cal J}_3^2.
\label{rr-2}
\end{equation}
If the state vector space on which Eq. (\ref{lr-cl}) is allowed to act is the
energy eigenspace, then the Hamiltonian $H$ in Eq. (\ref{lr-cl}) may be
replaced by the corresponding energy eigenvalue $E$, as the result,
Eq. (\ref{lr-cl}) can be put in the form of the Higgs algebra, Eq. (\ref{higgs}), with
\begin{equation}
    C_1 = \frac12 \lambda - 4 E ,
    \hspace{4mm}
    C_3 = 4 \lambda.
\label{ce-cs}
\end{equation}

Using Eqs. (\ref{casimir}), (\ref{lr-cl}) and (\ref{rr-2}), as expected, there
indeed exists a simple relation between $H$ and the Casimir invariant ${\cal C} $
of ${\cal H}$, i.e.
\begin{equation}
    H = 2 ( {\cal C} - \mu^2 ).
\label{ch-equ}
\end{equation}
It follows that calculation of the expectation value of Eq. (\ref{ch-equ})
in the Fock space ${\cal F}$,
with the help of Eq. (\ref{h-su2-casimir})
with setting $\alpha =0$ and Eq. (\ref{ce-cs}), leads immediately to the
following equation satisfied by $E$
\begin{equation}
    E = -2 \mu ^2 +  \left (- E +  \frac18 \lambda \right )
     N (N + 2 ) + \frac18 \lambda N^2 ( N + 2 )^2,
\label{k-e-equ}
\end{equation}
whose solution reads
\begin{equation}
    E_N = \frac{\lambda}{8} N (N+2) - \frac{2 \mu ^2 }{(N+1)^2}.
\label{k-e-solu}
\end{equation}
This result may also be obtained by using the Casimir invariant
(\ref{h-su11-casimir}) of ${\cal H}$ in the second kind of two-boson
realizations. Owing to the fact that $E_N$ depends only
to $N$ rather than $n_1$ and $n_2$, the degeneracy of the energy
level for the fixed $N$ is $N+1$. The physical condition that the
quantum number $\tilde{m} (= \frac12 (n_1- n_2))$ of ${\cal J}_3$
must be the non-negative integers requires that $N(= \frac12 (n_1 + n_2))$
has to take the non-negative even numbers, i.e., 0, 2, 4,....
If let $N=2n$ ($n=0$, 1, 2, ...), then Eq. (\ref{k-e-solu})
becomes the result (53) of Ref. \cite{higgs}. If the two
parameters $\lambda$ and $\mu$ in Eq. (\ref{k-e-solu}) satisfy the
following condition
\begin{equation}
    \frac{\mu ^2}{\lambda} = l \, \left (l+ \frac12 \right )^2 (l+1),
\end{equation}
where $l$ is some positive integer, then a zero energy level
appears at $N=2l$, i.e., $E_{2l} = 0$, while there exist $l$ bounded states,
$E_{2i} < 0$ ($i=0$, 1, ..., $l-1$), and infinite scattering states,
$E_{2j} > 0$ ($j=l+1$, $l+2$, ...).

\subsection{Phase operators of ${\cal H}$}
It is well known that the photon phase operators, introduced originally by Dirac
\cite{dirac} and amended by Susskind {\it et al} \cite{sg}, may be defined in terms of
one set of boson operator \{$a^+_1$, $a_1$, $\hat{n}_1$\} as \cite{carruthers}
\begin{eqnarray}
\begin{array}{rl}
    \exp (\mbox{i} \phi_1) = & \frac{1}{ \sqrt{ \hat{n}_1 + 1} } a_1,  \\
    \exp (- \mbox{i} \phi_1) = & a^+_1 \frac{1}{ \sqrt{ \hat{n}_1 + 1} }
         = (\exp (\mbox{i} \phi_1))^{\dagger}.
\end{array}
\label{ppo}
\end{eqnarray}
It is easily shown that the above two operators satisfy
\begin{eqnarray}
\begin{array}{rl}
    \exp (\mbox{i} \phi_1) |n_1 \rangle = & (1 - \delta_{n_1 0}) |n_1 -1 \rangle , \\
    \exp (- \mbox{i} \phi_1) |n_1 \rangle = & |n_1 +1 \rangle,
\end{array}
\label{ppo-state}
\end{eqnarray}
and
\begin{eqnarray}
\begin{array}{rl}
    (\exp (- \mbox{i} \phi_1))^{\dagger} \exp (- \mbox{i} \phi_1) = & 1 , \\
    \exp (- \mbox{i} \phi_1) (\exp (- \mbox{i} \phi_1))^{\dagger} = &
        1 - |0 \rangle \langle 0 |,
\end{array}
\label{semi-u}
\end{eqnarray}
hence, we call $\exp (\pm \mbox{i} \phi_1)$ semiunitary operators.
If introduce the following two Hermitian phase operators
\begin{eqnarray}
\begin{array}{rl}
    \cos \phi_1 = & \frac12 [\exp (\mbox{i} \phi_1) + \exp (- \mbox{i} \phi_1) ] , \\
    \sin \phi_1 = & \frac{1}{2\mbox{i}} [\exp (\mbox{i} \phi_1) - \exp (- \mbox{i} \phi_1) ],
\end{array}
\label{u}
\end{eqnarray}
then, they, together with $\hat{n}_1$, satisfy
\begin{eqnarray}
\begin{array}{l}
    [ \hat{n}_1, \: \cos \phi_1 ] = - \mbox{i} \sin \phi_1,  \cr
    [ \hat{n}_1, \: \sin \phi_1 ] = \mbox{i} \cos \phi_1.
\end{array}
\label{u-commu}
\end{eqnarray}

For the Higgs algebra ${\cal H}$, making use of the first kind of two-boson
realization, Eq. (\ref{su2-11-2}), we may construct the following two operators
\begin{eqnarray}
\begin{array}{rl}
    {\cal E}_+ = & \frac{2}{ \sqrt{ \hat{n}_1 + 1 } }
             \dot{B}^{(1,1)}_2 ( {\cal J}_- )  \frac{1}{ \sqrt{ (\hat{n}_2 + 1 )
            [ 2 C_1 + C_3 \hat{n}_1  (\hat{n}_2 + 1 ) ] } },     \\
    {\cal E}_- = & \frac{2}{ \sqrt{ ( \hat{n}_2 + 1)
        [ 2 C_1 + C_3 \hat{n}_1  (\hat{n}_2 + 1 ) ]} } \dot{B}^{(1,1)}_2 ( {\cal J}_+ )
        \frac{1}{ \sqrt{ \hat{n}_1 + 1 } } = ({\cal E}_+ )^{\dagger}.
\end{array}
\label{h-po}
\end{eqnarray}
We call ${\cal E}_{\pm}$ the phase operators of ${\cal H}$, since
action of ${\cal E}_{\pm}$ on the eigenvector $| \tilde{j} \tilde{m} \rangle$,
using Eqs. (\ref{amb-a}), (\ref{n12-am-b}), (\ref{urep}) and (\ref{su2-11-2}),
leads to
\begin{eqnarray}
\begin{array}{rl}
    {\cal E}_+ | \tilde{j} \tilde{m} \rangle
        = & (1 - \delta_{-\tilde{j} \tilde{m}}) | \tilde{j} \tilde{m}-1
        \rangle, \\
    {\cal E}_- | \tilde{j} \tilde{m} \rangle
        = & (1 - \delta_{\tilde{j} \tilde{m}}) | \tilde{j} \tilde{m}+1
        \rangle .
\end{array}
\label{e-act}
\end{eqnarray}
When $C_1=2$ and $C_3=0$, Eq. (\ref{h-po}) becomes the phase
operators of the angular momentum system. \cite{fan} Note that Eq.
(\ref{e-act}) is in fact the same as the equation satisfied by the
phase operators of the angular momentum system.

Using Eq. (\ref{ppo}), Eq. (\ref{h-po}) can also be written in the
following form
\begin{eqnarray}
\begin{array}{rl}
    {\cal E}_+ = & \exp [ \mbox{i} (\phi_1 - \phi_2)] w(\hat{n}_1,\hat{n}_2), \\
    {\cal E}_- = & w(\hat{n}_1,\hat{n}_2) \exp [- \mbox{i} (\phi_1 - \phi_2)]  ,
\end{array}
\label{h-pdo}
\end{eqnarray}
where $\exp [\pm \mbox{i} (\phi_1 - \phi_2)]$ are the ordinary phase difference operators
of two-dimensional harmonic oscillator, i.e.,
\begin{eqnarray}
\begin{array}{rl}
    \exp [ \mbox{i} (\phi_1 - \phi_2)] = & \frac{1}{\sqrt{ \hat{n}_1 +1}} a_1
        a^+_2 \frac{1}{\sqrt{ \hat{n}_2 +1}} , \\
    \exp [ - \mbox{i} (\phi_1 - \phi_2)] = & \frac{1}{\sqrt{ \hat{n}_2 +1}} a_2
        a^+_1 \frac{1}{\sqrt{ \hat{n}_1 +1}}   ,
\end{array}
\label{pdo}
\end{eqnarray}
and the operator function $w(\hat{n}_1,\hat{n}_2)$ is given by
\begin{eqnarray}
    w(\hat{n}_1,\hat{n}_2)  = \frac{2 \dot{f}^{(1,1)}_1 (\hat{n}_1 ,\hat{n}_2 ) }
            { \sqrt{ 2 C_1 + C_3 \hat{n}_1 (\hat{n}_2 + 1 ) }}
     =  \sqrt { \frac{ 4 C_1 +   C_3 [ \hat{n}_1^2
        + \hat{n}_2 (\hat{n}_2 + 2)] }
        { 4 C_1 + 2 C_3 \hat{n}_1 (\hat{n}_2 + 1 ) } },
\label{w-1}
\end{eqnarray}
where $\dot{f}^{(1,1)}_1 (\hat{n}_1 ,\hat{n}_2 )$ is the solution
of Eq. (\ref{solu-1}) with $\alpha = 0$ and $\dot{f}^{(1,1)}_1
(\hat{n}_1 ,\hat{n}_2 ) = \dot{g}^{(1,1)}_1 (\hat{n}_1 ,\hat{n}_2 )$.
Similar to the definition of nonlinear coherent state,
\cite{mmsz,mv} $w(\hat{n}_1,\hat{n}_2) \exp [- \mbox{i} (\phi_1 - \phi_2)]$
(see Eq. (\ref{h-pdo})) may be naturally called as the
nonlinear phase difference operator, which in fact plays the role of
amplifying the phase difference. Thus, Eq. (\ref{h-pdo})
shows that the phase properties of ${\cal H}$ can be described by
the nonlinear phase difference operator, while, as we know, the
phase properties of the angular momentum system may be described
by the phase difference operator of the two-dimensional harmonic
oscillator. \cite{fan}

Introduce another pair of Hermitian phase operators
\begin{eqnarray}
\begin{array}{rl}
    \cos \Phi = & \frac12 ({\cal E}_- + {\cal E}_+) , \\
    \sin \Phi = & \frac{1}{2\mbox{i}} ({\cal E}_- - {\cal E}_+),
\end{array}
\label{h-c-s}
\end{eqnarray}
it is easy to get
\begin{eqnarray}
\begin{array}{rl}
    [ {\cal J}_3, \: \cos \Phi ] = & - \mbox{i} \sin \Phi,  \cr
    [ {\cal J}_3, \: \sin \Phi ] = & \mbox{i} \cos \Phi ,
\end{array}
\label{h-jcs}
\end{eqnarray}
which is similar to Eq. (\ref{u-commu}).

\section{CONCLUSIONS}
In this paper we have obtained the explicit expressions for two
kinds of two-boson realizations of the Higgs algebra ${\cal H}$ by
generalizing the well known Jordan-Schwinger realizations of SU(2)
and SU(1,1). In each kind, the unitary realization, the
(constrained) nonunitary realizations of the $(1,1)$ case, and the
properties of their respective acting spaces have been discussed
in detail, together with the results of the $(2,2)$ case. The
other simple two-boson realizations for $k \not= l$, for example,
$(k,l)=(1,2)$, $(2,1)$, etc., have also been obtained by solving
Eq. (\ref{fg-su2}) and (\ref{fg-su11}), however, they are not
given here because of their complex expressions. It is worth
mentioning that for Eq. (\ref{h-equ}) in the first kind of
realizations, its solution (\ref{h-solu}), which can be found its
prototype for SU(2), is not unique, since, for example, it is
determined up to any periodic function ${\cal T}(m)$ of an
arbitrary but finite period $m$, namely, the constant $\alpha$ in
Eq. (\ref{h-solu}) may be replaced by ${\cal T}(m)$, and for the
$(1,1)$ case the general solution of Eq. (\ref{h-equ}) should be
$\hat{n}_1 + x(\hat{N})$, where $x(\hat{N})$ is an arbitrary
function of $\hat{N}(=\hat{n}_1 + \hat{n}_2)$. Similar properties
exist for Eq. (\ref{h-equ-2}) in the second kind of realizations.
Furthermore, we have revealed the fact that the nonunitary
realizations and the unitary ones may be related by the similarity
transformations, which have been obtained by solving the
corresponding unitarization equations satisfied by the nonunitary
realizations. Finally, as applications, first we have found that
the Kepler system in the two-dimensional curved space may be
described by the dynamical group chain, ${\cal H}$ $\supset$
$SO(2)$, that is, there exists a simple relation between the
Hamiltonian of this Kepler system and the Casimir operator of
${\cal H}$, and then obtained the energy levels by the eigenvalue
of the Casimir invariant. Secondly, we have constructed the phase
operators of the Higgs algebra in terms of the first kind of
two-boson unitary realization, which hold the similar properties
as the phase operators of the ordinary angular momentum systems.
Due to the tight relations between boson operators and
differential operators, for example, $a_i \leftrightarrow \partial
_{x_i}$ ($i=1$, 2) and $a_i^+ \leftrightarrow x_i$, the
two-variable differential realizations of the Higgs algebra may be
obtained directly from the above various two-boson realizations.
The method adopted in this paper may be naturally generalized to
the case of the multi-boson (or the deformed boson, the (deformed)
fermion, etc.) and be used to treat the general PAMA given by Eq.
(\ref{pama}).

\section*{ACKNOWLEDGMENTS}

This work is supported by National Natural Science Foundation of
China (10275038) and partly by 
Major State Basic Research Development Programs
(G2000077400 and G2000077604) and Tsinghua Natural Science
Foundation (985 Program).


\vspace{.5cm}

\begin{thebibliography}{99}
\bibitem{rocek} M. Ro\v{c}ek, Phys. Lett. B {\bf 255}, 554 (1991).
\bibitem{bl} L.C. Biedenharn and J.D. Louck,
    {\it Angular Momentum in Quantum Physics}
    (Addison-Wesley, Massachusetts, 1981).
\bibitem{higgs} P.W. Higgs, J. Phys. A: Math. Gen. {\bf 12}, 309 (1979).
\bibitem{zhedanov} A.S. Zhedanov, Mod. Phys. Lett. A {\bf 7}, 507 (1992).
\bibitem{jimbo} M. Jimbo, Lett. Math. Phys. {\bf 10}, 63 (1985).
\bibitem{dask} C. Daskaloyannis, J. Phys. A: Math. Gen. {\bf 24}, L789 (1991).
\bibitem{bdk} D. Bonatsos, C. Daskaloyannis, and K. Kokkotas,
              Phys. Rev. A {\bf 50}, 3700 (1994).
\bibitem{bkd} D. Bonatsos, P. Kolokotronis, and C. Daskaloyannis,
              Mod. Phys. Lett. A {\bf 10}, 2197 (1995).
\bibitem{quesne1} C. Quesne, Phys. Lett. A {\bf 193}, 245 (1994).
\bibitem{jr} G. Junker and P. Roy, Phys. Lett. A {\bf 257}, 113
(1999).
\bibitem{sbjps} V. Sunilkumar, B.A. Bambah, R. Jagannathan, P.K.
Panigrah, and V. Srinivasan, J. Opt. B {\bf 2}, 126 (2000).
\bibitem{sbj} V. Sunilkumar, B.A. Bambah, and R. Jagannathan,
arXiv:math-ph/0205005 (2002).
\bibitem{bbd} J. Beckers, Y. Brihaye, and N. Debergh, J. Phys. A: Math. Gen. {\bf 32},
       2791 (1999).
\bibitem{debergh-1} N. Debergh, J. Phys. A: Math. Gen. {\bf 31}, 4013 (1998).
\bibitem{debergh-2} N. Debergh, J. Phys. A: Math. Gen. {\bf 33}, 7109 (2000).
\bibitem{rjr} D. Ruan, Y.F. Jia, and W. Ruan, J. Math. Phys. {\bf 42}, 2718
(2001).
\bibitem{schwinger} J. Schwinger, {\it The Quantum Physics of Angular Momentum},
    edited by L.C. Biedenharn and H. Van Dam (Academic, New York, 1965).
\bibitem{km} A. Klein and E.R. Marshalek, Rev. Mod. Phys. {\bf 63},
375 (1991).
\bibitem{fradkin} D.M. Fradkin, J. Phys. A: Math. Gen. {\bf 27},
1261 (1994).
\bibitem{ruan} D. Ruan, {\it Frontiers in Quantum Mechanics},
edited by J.Y. Zeng, S.Y. Pei, and G.L. Long (Beijing University,
Beijing, 2001).
\bibitem{hp-1} T. Holstein and H. Primakoff, Phys. Rev. {\bf 58},
1098 (1940).
\bibitem{dyson} J.F. Dyson, Phys. Rev. {\bf 102}, 1217 (1956).
\bibitem{ruan1} D. Ruan, to be published in Phys. Lett. A. (2003).
\bibitem{iachello} F. Iachello, Rev. Mod. Phys. {\bf 65}, 569
(1993).
\bibitem{pauli} W. Pauli, Z. Phys. {\bf 36}, 336 (1926).
\bibitem{fock} V.A. Fock, Z. Phys. {\bf 98}, 145 (1935).
\bibitem{bargmann} V. Bargmann, Z. Phys. {\bf 99}, 576 (1936).
\bibitem{dirac} P.A.M. Dirac, Proc. Roy. Soc. A (London) {\bf 114}, 243 (1923).
\bibitem{sg} L. Susskind and J. Glogower, Physics {\bf 1}, 49 (1964).
\bibitem{carruthers} P. Carruthers and M.M. Nieto, Rev. Mod. Phys. {\bf 40}, 411 (1968).
\bibitem{fan} H.Y. Fan and Y.P. Li, Commun. Theor. Phys. {\bf 9},
341 (1988).
\bibitem{mmsz} V.I. Man'ko, G. Marmo, E.C.G. Sudarshan, and F.
Zaccaria, Phys. Scr. {\bf 55}, 528 (1997).
\bibitem{mv} R.L. de Matos and W. Vogel, Phys. Rev. A {\bf 54},
4560 (1996).
\end{thebibliography}
\end{document}